\documentclass[final,printer]{rQUF2e}

\usepackage{epstopdf}
\usepackage{subfigure}
\usepackage{tabularx} 
\usepackage{booktabs} 
\usepackage{multirow} 
\usepackage{tablefootnote}
\usepackage{lmodern}
\usepackage{svg}

\usepackage{url}

\newcolumntype{P}[1]{>{\centering\arraybackslash}p{#1}}

\theoremstyle{plain}

\theoremstyle{definition}

\theoremstyle{remark}

\makeatletter
\def\blfootnote{\xdef\@thefnmark{}\@footnotetext}
\makeatother

\begin{document}


\title{Stock Movement Prediction with Financial News using Contextualized Embedding from BERT}

\author{QINKAI CHEN$^{\ast}$\thanks{$^{\ast}$Corresponding Author. Email: qinkai.chen@polytechnique.edu.
The author would like to thank Christian-Yann Robert from ENSAE Paris and Mathieu Rosenbaum from 
Ecole Polytechnique for their valuable guidance and advice in this research.
The author also gratefully acknowledges the help of Jean-Sebastien Deharo and Alexandre Davroux from Exoduspoint Capital Management France in this work.
}
\affil{Ecole Polytechnique, Route de Saclay, 91120 Palaiseau, France \\
Exoduspoint Capital Management France, 32 Boulevard Haussmann, 75009 Paris, France
}
\received{\today}
}

\maketitle

\begin{abstract}
  News events can greatly influence equity markets.
  In this paper, we are interested in predicting the short-term movement of stock prices after financial news events 
  using only the headlines of the news.
  To achieve this goal, we introduce a new text mining method called Fine-Tuned Contextualized-Embedding
  Recurrent Neural Network (FT-CE-RNN). 
  Compared with previous approaches which use static vector representations of the news (static embedding),
  our model uses contextualized vector representations of the headlines (contextualized embeddings) generated from
  Bidirectional Encoder Representations from Transformers (BERT).
  Our model obtains the state-of-the-art result on this stock movement prediction task.
  It shows significant improvement compared with other
  baseline models, in both accuracy and trading simulations.
  Through various trading simulations based on millions of headlines from Bloomberg News,
  we demonstrate the ability of this model in real scenarios.
\end{abstract}


\begin{classcode}C67, G11, G14\end{classcode}

\newpage
\section{Introduction}
\label{sec:introduction}

Stock movement prediction has attracted a considerable amount of attention
since the beginning of the financial market, although the stock prices are
highly volatile and non-stationary.
\citet{fama1965behavior} showed that the movement of stock prices can be
explained jointly by all known information.

\begin{figure}[h]
  \centering
  \includegraphics[width=\columnwidth]{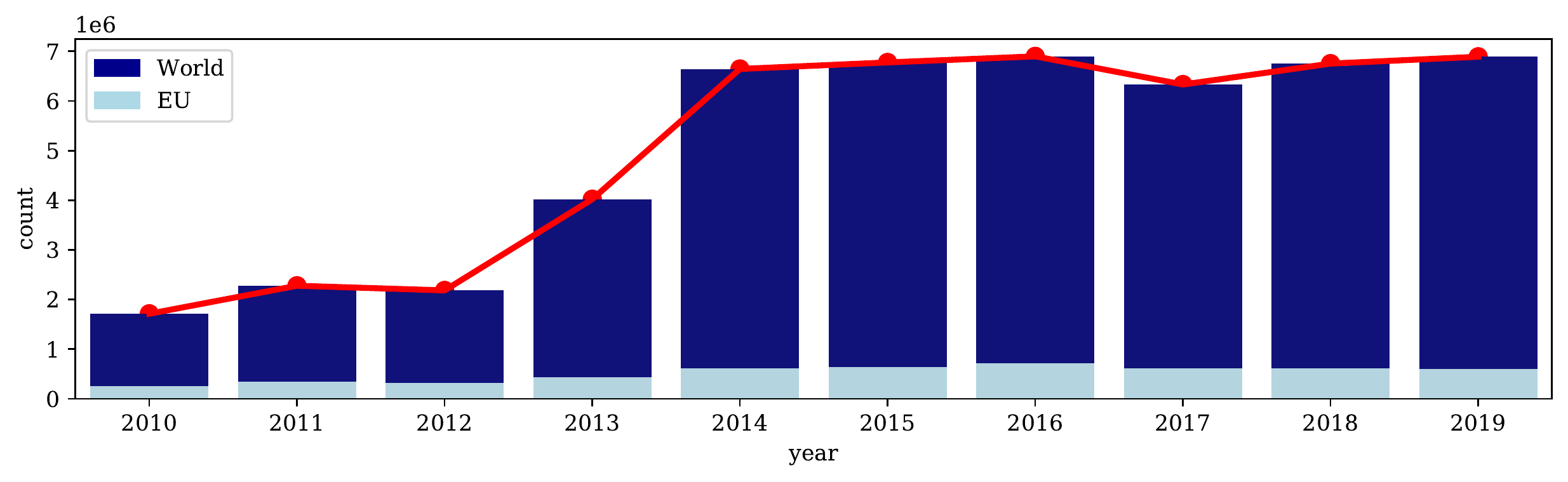}
  \caption{The number of news recorded by Bloomberg each year.
  For both world and European countries, there is
  a significant increase in the number of news from 2014.}
  \label{fig:news_count}
\end{figure}

With the development of Internet, there is a rapid increase in the amount of financial news
data (Figure \ref{fig:news_count}), and more studies have been done to use computational methods to predict
stock price changes based on financial news
\citep{oliveira2013some,si2013exploiting,xie2013semantic,nguyen2015topic,luss2015predicting,rekabsaz2017volatility,ke2019predicting,li2020identifying,coqueret2020stock}.
Following previous works, we explore an accurate method to transform 
textual information into stock movement prediction signal.

\citet{schumaker2009textual} use a classical feature engineering method to
predict the market behavior. More recently, deep learning methods are more
frequently applied on this task. \citet{ding2014using, ding2015deep} employ
structured representations to normalize a news then apply a Convolutional 
Neural Network (CNN) on this formulation. 
\citet{hu2018listening} apply an improved Transformer model \citep{vaswani2017attention} 
to handle all the words in the raw text simultaneously to predict the forward return.
\citet{luss2015predicting} propose a statistical learning method to 
combine text data and the historical returns.
\citet{xu2018stock} improve the idea from \citet{luss2015predicting} by 
designing a deep neural network. 
\citet{ke2019predicting} 
adopts a simple but effective classification method combining both regression  
and Term-Frequency Inversed Document Frequency (TF-IDF) model \citep{jones1972statistical}. 
\citet{del2020unsupervised} further introduce 
an unsupervised method to extract market moving events from text data, which overcomes
the problem of lacking reliable labels in financial data.

Before applying computational models mentioned above, the first step usually involves
converting words into fixed-length vectors (these fixed-length vectors are called embeddings
in natural language processing, details are presented in Section \ref{subsec:contextualized_embedding}). 
\citet{mikolov2013distributed} propose Word2Vec model
to embed words based on words co-occurrence prediction and \citet{pennington2014glove} propose
a similar GloVe model based on words co-occurrence frequencies. However, both methods can only
generate static non-contextualized embeddings. It means that a word is converted to the same
vector no matter its meaning or its context.
This approach ignores the fact that the meaning of a word can change significantly
in different contexts, which impacts the performance of the model.
As most of the previous researches rely heavily on static non-contextualized
embedding such as Word2Vec or GloVe, there can be accuracy loss.

\citet{cer2018universal} and \citet{peters2018deep} propose methods to generate contextualized
embeddings by jointly considering all the words together. They published their models 
trained on large English corpus. Although effective, the model is fixed and does not contain 
domain-specific knowledge in finance.

\citet{devlin2018bert} introduce a general-use language model called Bidirectional Encoder Representations from Transformers (BERT). 
It is one of the most promising models in natural language processing and it 
showed significant improvement on multiple benchmarks \citep{wang2018glue}.
The BERT model is pre-trained on a very large scale of textual data to leverage all the 
features in natural languages, and it also provides the ability to fine-tune this pre-trained
model with domain-specific data without needing to start from the scratch. As we have 
a large amount of financial texts, we can use them to add financial knowledge to the BERT model 
and generate contextualized embeddings with domain-specific knowledge in finance from BERT.


In addition, previous researches evaluate the performance of the models based 
on the accuracy calculated on all the news \citep{ding2015deep,xu2018stock,hu2018listening}. 
However, this evaluation metric does not reflect the real capability of the model since 
investors only care about the news which can move the market significantly.
The news identified as neutral have little impact on investors' decisions, 
as investors will simply ignore the news if they are classified as neutral.

In this paper, aiming to solve the problems mentioned above, we want our research to have 
the following characteristics:
\begin{itemize}
  \item It adopts the contextualized embeddings instead of the static embeddings.
  \item The contextualized embeddings contain financial domain-specific knowledge.
  \item Our model has a better prediction on the news which can move the market significantly.
\end{itemize}

Hence, based on previous work (Sec. \ref{sec:related_work}),
we propose Fine-Tuned Contextualized-Embedding Recurrent Neural Network (FT-CE-RNN)
to predict the stock price movement based on the headlines (Sec. \ref{sec:problem_formulation}). 
Using Bloomberg News dataset (Sec. \ref{sec:data}), 
this model generates contextualized embeddings with domain-specific knowledge using all the hidden vectors 
from the BERT model fine-tuned on financial news. Then FT-CE-RNN uses a recurrent neural network (RNN)
to make use of the generated embeddings. (Sec. \ref{sec:prediction_model})
We also introduce a new evaluation metric
which calculates the accuracy on various percentiles of the prediction
scores on the test set instead of the whole test set to better incorporate investors' interests.
Our experiments show that our FT-CE-RNN achieves a state-of-the-art
performance compared with other baseline models. 
We also evaluate our model by running trading simulations with different trading strategies. (Sec. \ref{sec:experiments})

\section{Related Work}
\label{sec:related_work}

\subsection{Stock Movement Prediction}

Stock movement prediction is a widely discussed topic in both finance and computer science communities.
Researchers predict the stock market using all available information, including
historical stock price, company fundamentals, third-party news sentiment score, financial news, social media texts
and even satellite images.

The most classical method is to use 
the historical stock prices to predict the future prices. 
\citet{kraft1977determinants,sonsino2014return,ariyo2014stock,kroujiline2016forecasting,jiang2018short}
use time series analysis techniques to extract the patterns of historical returns,
and predict the future stock movement based on these patterns.
More recently, researchers start to use neural networks to analyze this pattern
\citep{kohara1997stock,adebiyi2012stock,tashiro2019encoding,chen2019exploring,makinen2019forecasting,bai2020machine}.

Financial analysts usually use companies' fundamental
  indicators from their financial reports to predict the stocks' prices in the future \citep{zhang2018modelling}.
  This includes the use of earnings per share (EPS) \citep{patell1976corporate},
  debt-to-equity (D/E) ratio \citep{bhandari1988debt}, cash flow \citep{liu2007cash}, etc.
  \citet{nonejad2021bayesian} builds a conditional model to jointly consider historical prices
  and financial indicators.

With the rapid development of the natural language processing and deep learning, 
researchers start to focus on predicting stock movement based on textual data, such as
financial news and social media texts, which were viewed as difficult to process systematically.
Financial news data vendors such as Bloomberg, ThomsonReuters and RavenPack all include their proprietary
sentiment analysis on the news. 
\citet{coqueret2020stock} thoroughly analyzes the sentiment classification given by Bloomberg and finds 
disappointing results on its predicting power. \citet{ke2019predicting} include the RavenPack's proprietary
score as a benchmark and find it less performing than other models.

Hence, more researchers propose their own natural language processing models to improve the predictability based on financial news.
\citet{luss2015predicting} propose an improved Kernel learning method to extract the features in the texts.
\citet{ke2019predicting} use statistical learning methods to determine the sentiment of 
the words in the news.
More recently, computer scientists begin using state-of-the-art deep learning techniques
to solve this problem. \citet{ding2015deep,hu2018listening,xu2018stock,li2020identifying} propose
different deep learning models to extract information from both financial news and 
social media texts.

There are also other researches which use uncommon data to predict future stock prices. 
The data includes key people compensation \citep{cooper2016performance}, 
satellite images \citep{donaldson2016view} and the pictures included in the news 
\citep{obaid2021picture}.

\subsection{Contextualized Embedding}
\label{subsec:contextualized_embedding}

In the natural language processing, the first step usually involves transforming words or sentences
into fixed-length vectors to allow numerical computations. These fixed-length vectors are known as
the embeddings of the words or the sentences.

Historically, researchers use one-hot embeddings \citep{stevens1946theory} to encode words. 
However, the dimension of the one-hot embedding is large since each unique word takes 
one dimension. Hence, researchers start to develop methods to make the word embeddings denser.

The most widely used methods for word embedding are Word2Vec \citep{mikolov2013distributed} and
GloVe \citep{pennington2014glove}, both of which are based on word co-occurrences. Such models 
take in a large number of texts and output a fixed vector for each word in the texts.
The more frequently the two words co-occur, the more correlated two embeddings are.
Once the model is trained, the embeddings of the words no long change, therefore we call
these embeddings static embeddings. Such model generates the same embedding for one 
word no matter its context, although the meanings of the words can depend on the context 
in which this word occurs.

Researchers propose contextualized embeddings to solve this issue. Instead of taking only
one words as input, the contextualized embedding model accepts the whole sentence as its input.
The model then generates the embeddings for each word in the sentence by jointly considering
the word and all the other words in the sentence. \citet{cer2018universal} proposes Universal 
Sentence Encoder (USE) to encode the whole sentence contextually. However, USE only gives 
the embedding of the sentence as a vector without specifying the embedding of each word.
\citet{peters2018deep} proposed Embeddings from Language Models (ELMo) to embed words 
based on their linguistic contexts, but ELMo is trained on general English language,
making the generated embeddings lack of financial domain-specific knowledge.
However, \citet{yang2020finbert}
showed domain-specific model outperforms general models in most of the tasks.

Recent researches on general-use language model such as BERT 
\citep{devlin2018bert} and XLNet \citep{yang2019xlnet} reported impressive result on all
natural language processing tasks. More interestingly, these models propose a way to fine-tune
its pre-trained model on general English with domain-specific data. 

Hence, we propose FT-CE-RNN to complement existing researches. FT-CE-RNN generates 
contextualized embeddings with domain-specific knowledge from the BERT model,
it then makes the stock movement prediction based on this more advanced embedding.

\section{Problem Formulation}
\label{sec:problem_formulation}

Suppose that we have a stock $s$ with a headline $h_{s,t}$ recorded at time $t$,
and the headline has $N$ words, we denote them by $w_{1}, ..., w_{N}$.
We first need to transform them into fixed-length embeddings. Suppose that the length
of the embedding is $l_{e}$, this process can be written as:
\begin{equation}
  Emb_{i} = f_{s}(w_{i})
\end{equation}
where $Emb_{i} \in \mathbb{R}^{l_{e}}$ is the embedding of the word $w_{i}$
and $f_{s}$ denotes the static embedding encoder. In this case, each word has a fixed 
embedding independent of its context.

A contextualized embedding encoder has the same function of converting a word into 
a vector, unless it considers all the words in a sentence together. We use
$f_{c}$ to denote this contextualized embedding encoder, it can be written as:
\begin{equation}
  Emb_{i} = f_{c}(w_{i}|w_{1}, ..., w_{N})
\end{equation}

We concatenate the embeddings of all words to get the embedding of the headline 
$h_{s,t}$. We define the embedding of this headline as:
\begin{equation}
  Emb_{h_{s,t}} = [Emb_{1}, ..., Emb_{N}]
\end{equation}

\noindent
where $Emb_{h_{s,t}} \in \mathbb{R}^{l_{e} \times N}$.

Following the work of \citet{luss2015predicting,ding2015deep,xu2018stock,ke2019predicting},
we formulate the stock movement prediction as a binary classification task\footnote{
  We can also formulate this problem as a multi-class classification task \citep{pagolu2016sentiment}.
  It means that, instead of classifying a news into positive news and negative news, we can 
  classify them into positive news, negative news or neutral news, making it a three-class
  classification task. Moreover, we can classify a news into different return intervals, making it
  a multi-class classification task.
However, we find that the performance with multi-class classification
setup is less impressive. We provide the details of this study in Section \ref{subsec:multi_class_result}.}.
It means that we predict if a news has a \textbf{positive} impact or a \textbf{negative} impact on the related stock.

We define its market-adjusted return $r_{s, t}$ as
\begin{equation}
  \label{eq:market_adjusted_return}
    r_{s,t} = \frac{P_{s, t + \Delta t}}{P_{s, t}} - \frac{P_{m, t + \Delta t}}{P_{m, t}}
\end{equation}
where $P_{s, t}$ denotes the price of stock $s$ at time $t$ and 
$P_{m, t}$ denotes the value of the equity index at time $t$.

We notice that it is necessary to use market-adjusted return instead of
the simple return, as the information contained in the price change is
partially due to the information related to this stock (such as news), and also
partially due to the information related to other macroeconomic information (such
as interest rate, fiscal policies, etc.). As the macroeconomic effect impacts all
stocks, it can be explained by a weighted sum of all stocks, such as market index.
We can simply remove this impact by subtracting the index return from the
stock return, and this adjusted return can better explain the impact
of the news.

Most researches in the stock movement prediction based on news simply suppose that all the 
news induce the market change in the same way \citep{luss2015predicting,xu2018stock,hu2018listening,ke2019predicting}, and 
therefore use the same $\Delta t$ to calculate the forward returns of all news. However,
\citet{fedyk2018front} suggests that the news during the trading hours and the news outside
the trading hours have different market impact.

Hence, for different news, we choose different $\Delta t$. 
For example, for the news published during the trading hours, 
the price can change in several minutes after the arrival of the news. 
In this case, we can choose a smaller $\Delta t$ of several minutes or several hours. 
However, for the news published out of 
the trading hours, as the market is already closed, we cannot observe the effect of the news
until the next market open. Therefore, we need to choose a $\Delta t$ of several days.

We define the stock price movement as:
\begin{equation}
  \label{eq:problem_formulation}
  Y_{s, t} =
  \begin{cases}
    1 , & r_{s, t} > 0\\
    0 , & r_{s, t} \leq 0\\
  \end{cases}
\end{equation}

The goal is to predict $Y_{s,t}$ from the embeddings of the headlines $Emb_{h_{s,t}}$. It can be written as:
\begin{equation}
  \hat{Y}_{s,t} = g(Emb_{h_{s,t}})
\end{equation}
where $g$ represents the prediction model.

\section{Data}
\label{sec:data}

\subsection{Data Description}
\label{subsec:data_description}

The dataset that we use is Bloomberg News\footnote{https://www.bloomberg.com/professional/product/event-driven-feeds/}.
In this dataset, each entry contains a \textit{timestamp} showing when this news is published, a \textit{ticker} which 
tells the stock related to this news and the \textit{headline} of this news. In addition to the necessary information
above, there are two fields given by Bloomberg's proprietary classification algorithm.
The \textit{score} is among -1, 0 and +1, which indicates if the news is either negative, neutral or positive.
\textit{Confidence} is a value between 0 and 100 related to \textit{score}.
A higher \textit{confidence} value means that Bloomberg's model is more sure about its \textit{score}.
Bloomberg's classification will serve as one of the benchmarks for our prediction model. 
We present a sample dataset in Table \ref{tab:data_sample}.

\begin{table}[h]
  \begin{center}    
  \tbl{A small sample from the Bloomberg News dataset}
  {
    \begin{tabular}{P{20em}P{10em}ccc}
    \toprule
    \textbf{Headline} & \textbf{TimeStamp} & \textbf{Ticker} & \textbf{Score} & \textbf{Confidence} \\
    \midrule
    1st Source Corp: 06/20/2015 - 1st Source announces the promotion of Kim Richardson in St. Joseph & 2015-06-20T05:02:04.063 & SRCE & -1 & 39 \\
    \midrule
    Siasat Daily: Microsoft continues rebranding of Nokia Priority stores in India opens one in Chennai & 2015-06-20T05:14:01.096 & MSFT & 1 & 98 \\
    \midrule
    Rosneft, Eurochem to cooperate on monetization at east urengoy & 2015-06-20T08:01:53.625 & ROSN RM & 0 & 98 \\
    \bottomrule
    \end{tabular}}
  \label{tab:data_sample}%
\end{center}
\end{table}%

We need to address that in our dataset, we only have the headlines of the news instead of the whole article.

In our experiment, we use the news data on all the stocks from the STOXX Europe 600 index\footnote{https://www.stoxx.com/index-details?symbol=SXXP}
which represents the 600 largest stocks of the European market. In order not to overfit, we select a short period 
(from 01/01/2016 to 30/06/2018) as our training set and another short period (from 01/07/2018 to 31/12/2018) as our development set.
We tune the parameters of the models only based on this subset of the data, we then test on the whole period (from 01/01/2011 to 31/12/2019) 
on a 3-year rolling basis. It means that we generate the classification result of the year $y$ using model trained between $y-3$ and $y-1$, 
without varying the parameters initially obtained. Detailed statistics of the dataset are in Table \ref{tab:data_stats}.
We can also find the number of news in each year from Figure \ref{fig:news_count}.

\begin{table}[h]
  \begin{center}
  \tbl{Statistics of the Bloomberg News dataset}
  {
    \begin{tabular}{cccc}
    \toprule
      & \textbf{Train} & \textbf{Dev} & \textbf{Test}$^{\rm a}$ \\
    \midrule
    Total news & 1,616,922 & 316,944 & 5,253,345 \\
    Word counts & 17,650,629 & 3,554,324 & 55,410,309 \\
    From date & 01/01/2016 & 01/07/2018 & 01/01/2011 \\
    End date & 30/06/2018 & 31/12/2018 & 31/12/2019 \\
    \bottomrule
    \end{tabular}}
    \label{tab:data_stats}%
  \end{center}
    \tabnote{$^{\rm a}$ Note that we do not simply apply our model trained on the training set on the test set. 
    The training set and the development set are only used to find the hyper-parameters for our models. 
    We generate the scores on test set using different models trained on a 3-year rolling basis, as described in section \ref{subsec:data_description}}
\end{table}%

In addition to the Bloomberg news dataset, we also use the cooperate action adjusted share prices 
at market close and intraday minute bar share prices for all the stocks. 
The share prices are used to label our data and simulate our trading strategies.

\subsection{Data Labelling}
\label{subsec:data_labelling}

For this supervised learning task, we need to provide our model with the ground-truth as its target.
However, our data is simply the headlines of the financial news,
it does not tell us if a news is positive or negative. Hence, we need to give each news in the training set 
a label (positive or negative) before training our model.

\begin{figure}[h]
  \centering
  \subfigure[In trading hours]{\includegraphics[width=0.45\linewidth]{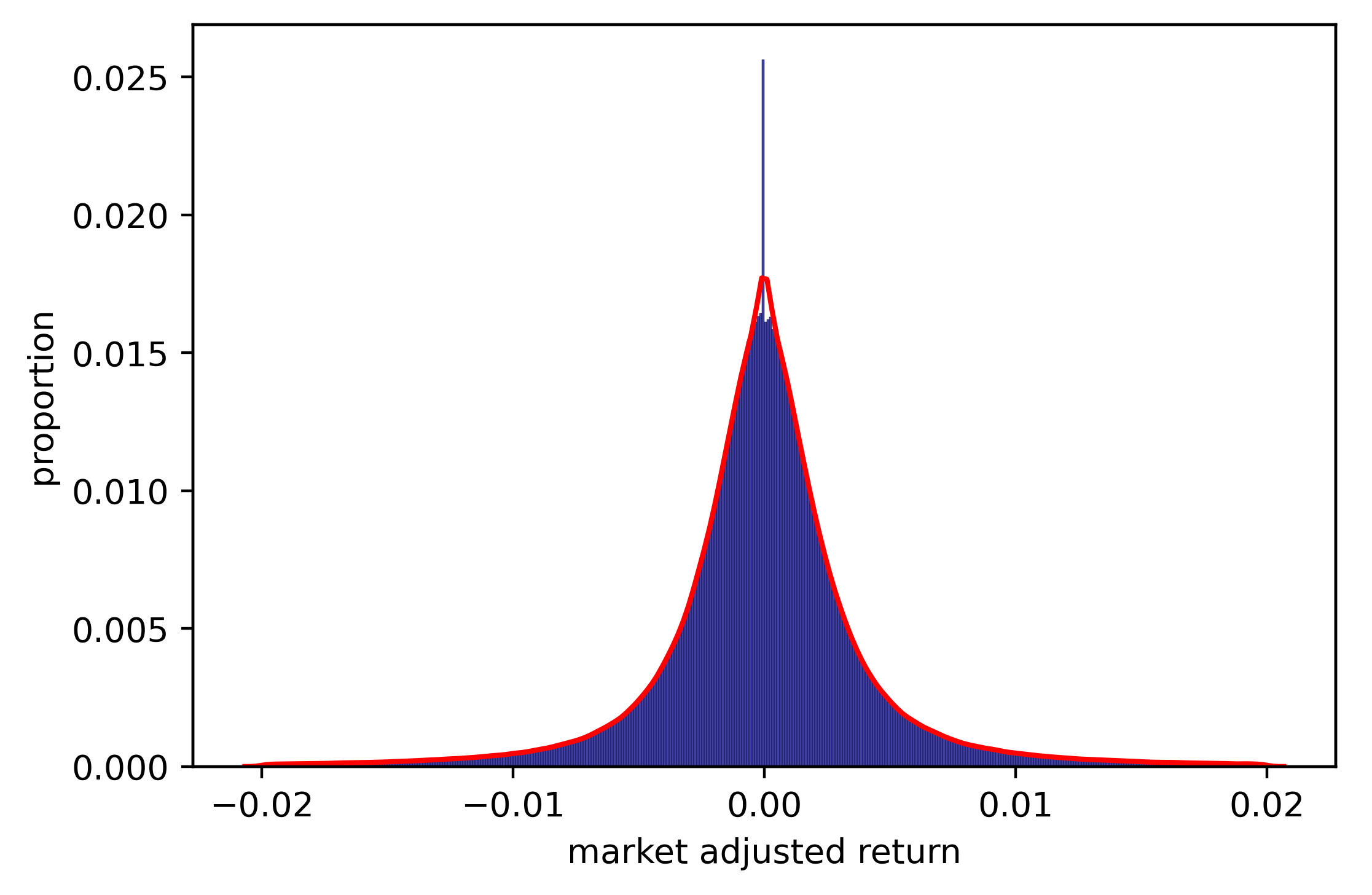}}
  \subfigure[Outside trading hours]{\includegraphics[width=0.45\linewidth]{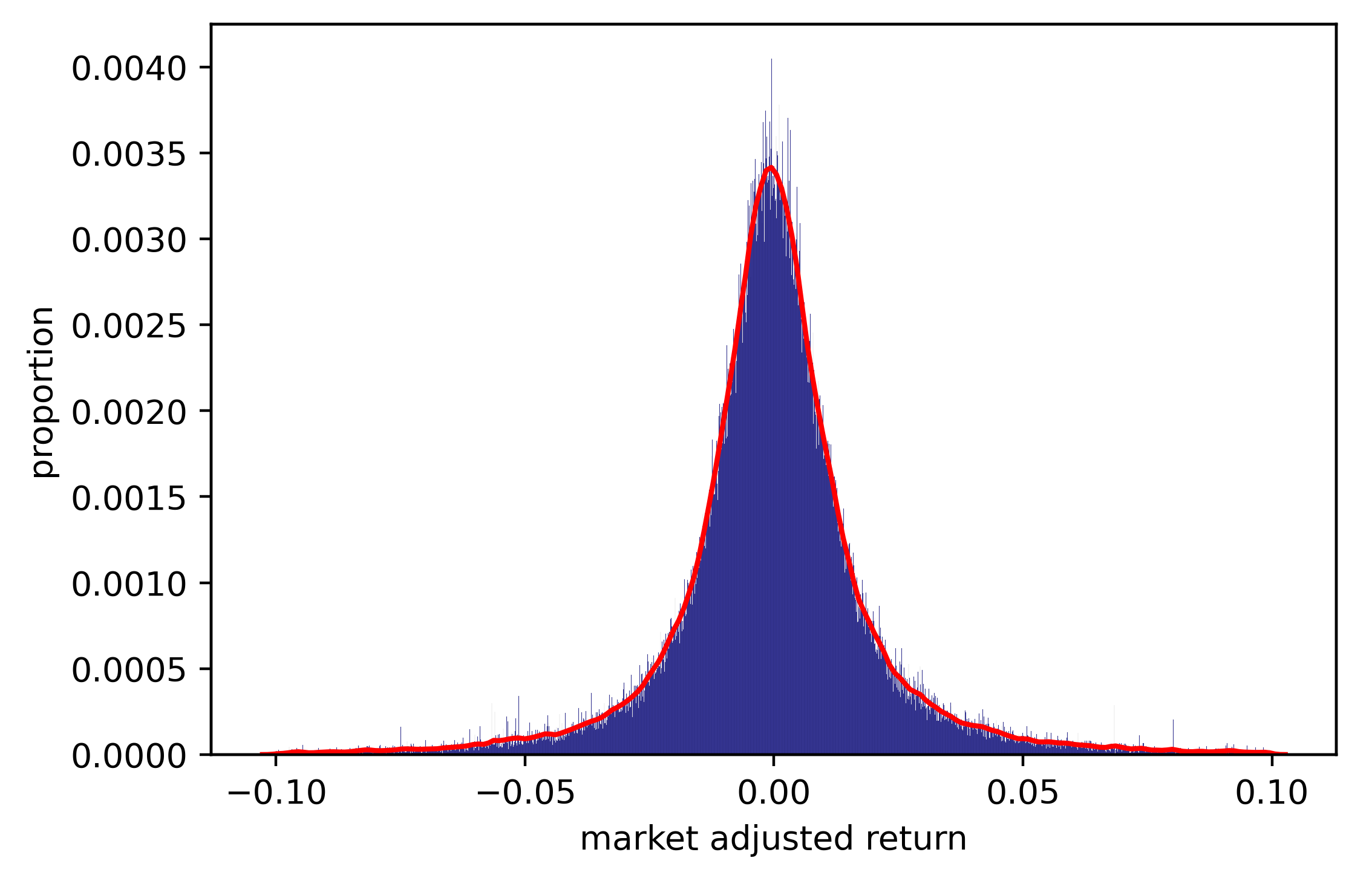}}
  \caption{The distribution of the $r_{s,t}$ for both the news
  in the trading hour and those outside the trading hour.
  The returns are calculated on all the news in the training set.
  For the news in the trading hour (Figure (a)), we use the forward return of 30 minutes ($\Delta t_{i} = 30 minutes$), 
  and for those outside the trading hour (Figure (b)), we use the forward return of 1 day($\Delta t_{e} = 1 day$).
  We can see that the market adjusted returns are symmetrically distributed.
  }%
  \label{fig:distribution_return}
\end{figure}

The intuition behind our labelling method is simply that if a news is positive, the investors will 
start to overbuy the related stocks, 
the stock should therefore outperform the market and vice versa. 
We use Equation \ref{eq:market_adjusted_return} to calculate the market adjusted return for each news.
As discussed in Section \ref{sec:problem_formulation}, we consider the different 
effects of the news which occur during the trading hour\footnote{9:00 to 17:30 CET for most European markets} 
and those outside the trading hour.
We use different $\Delta t$ (Eq. \ref{eq:market_adjusted_return}) for the news during the trading hours and 
outside the trading hours. For the news inside trading hours, we choose $\Delta t_{i}$ 
and for the news outside the trading hours, we use $\Delta t_{e}$. \footnote{For $\Delta t_{i}$, we test 5,30,60,120 minutes; 
for $\Delta t_{e}$, we test 1,2,3,5 days.}
We show an example of the distribution of the returns for all the news in Figure \ref{fig:distribution_return}.

We label the data based on the market-adjusted return mentioned above. As our task
is to identify market-moving news which investors focus on, we need to remove news which do not
have significant impact on the price. As we found that the return distribution is
quite balanced for the training set (Figure \ref{fig:distribution_return}), 
we simply label the 15\% news with most positive
return as 1 and the 15\% news with most negative return as 0. 
This can be written as:
\begin{equation}
  Y_{s, t, train} =
  \begin{cases}
    1 , & r_{s, t}\ in\ top\ 15\%\\
    0 , & r_{s, t}\ in\ bottom\ 15\%\\
    removed , & otherwise \\
  \end{cases}
\end{equation}
where $Y_{s, t, train}$ is the label for the news $e_{s,t}$ for stock $s$ recorded at time $t$
in the training set.

However, for development and test sets, 
we label \textbf{all} news with positive return as 1 and all news with negative
return as 0. This can be written as:
\begin{equation}
  Y_{s, t, dev/test} =
  \begin{cases}
    1 , & r_{s, t} > 0\\
    0 , & r_{s, t} \leq 0\\
  \end{cases}
\end{equation}
This difference in labelling is simply to avoid the information leakage. 
In real-life scenario we cannot know the forward return of a news when it is published. 
Therefore, we cannot know if the news is in the top 15-percentile or the bottom 15-percentile.
We are supposed to give each news a score when it is published regardless its 
forward return.

However, we can choose to exclude a news according to its score when calculating the metrics,
as this information is available immediately after we receive the news. We use this idea 
to construct different test sets to evaluate of model. We present the details in Section
\ref{subsec:evaluation_metrics}.

\section{Prediction Model}
\label{sec:prediction_model}

There are two main components in our model: a contextualized embedding encoder
from BERT model and a Recurrent Neural Network (RNN) which takes
the contextualized embedding as input and outputs the classification probability for both classes.

\subsection{Contextualized Embedding Encoder from BERT}
\label{sec:ce}

The Bidirectional Encoder Representations from Transformers (BERT) proposed by 
\citet{devlin2018bert} is a widely used language model in the natural language 
processing applications. It is first pre-trained on very large scale data 
(WikiBooks\footnote{https://en.wikibooks.org/}
and Wikipedia\footnote{https://en.wikipedia.org/}) to learn the basic characteristics
of a language. After this pre-training phase, we can obtain a pre-trained base BERT model 
for all other downstream tasks (such as text classification). This pre-training process is computationally intensive,
we simply use the pre-trained BERT model published by 
Google\footnote{https://storage.googleapis.com/bert\_models/2020\_02\_20/uncased\_L-12\_H-768\_A-12.zip}.

\begin{figure}[h]
  \centering
  \includegraphics[width=0.8\columnwidth]{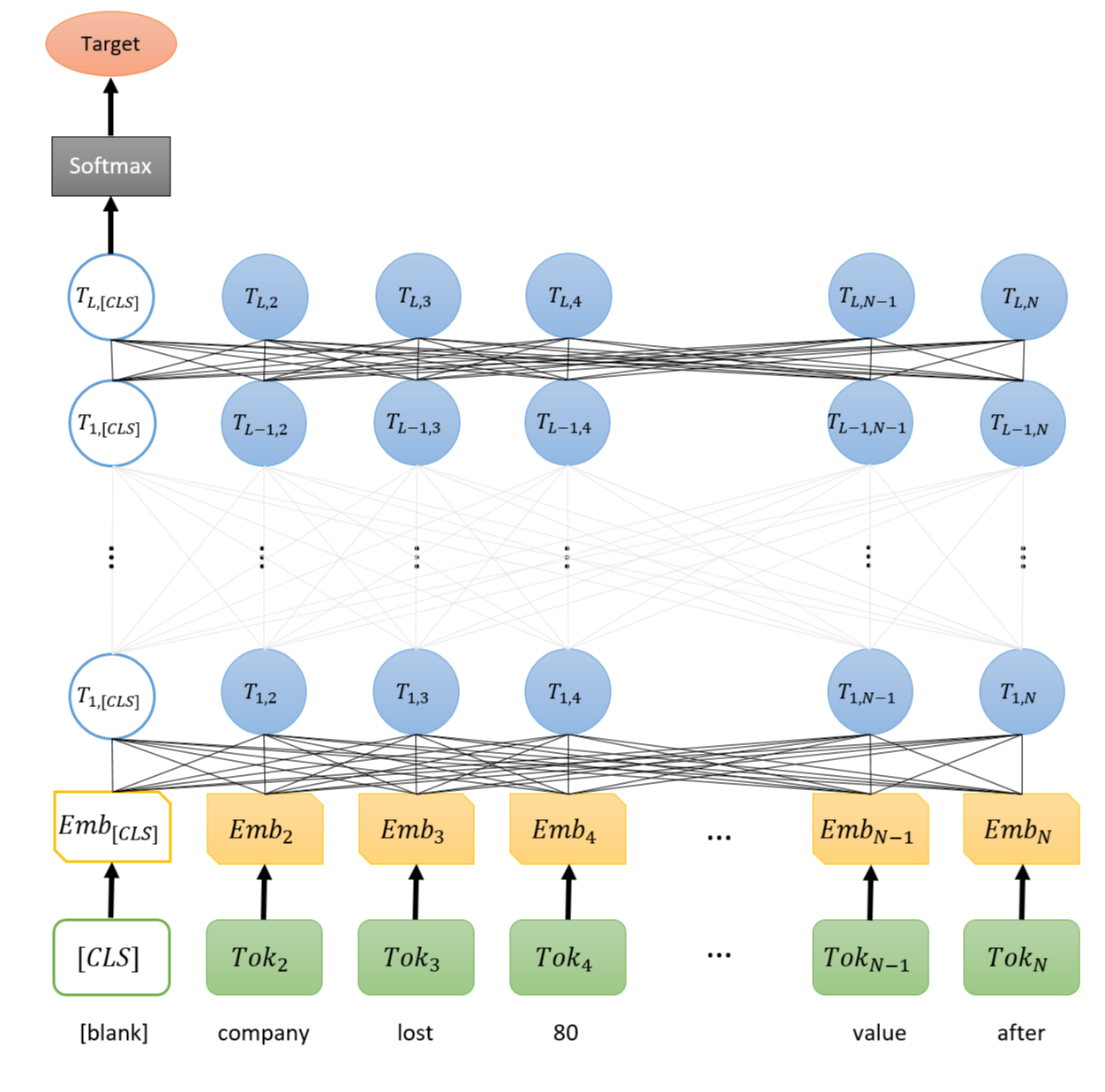}
  \caption{BERT model with input headline
    "\textit{company lost about 80 percent of its market value after interim data}" and the maximum length of model is set to 10 ($N=10$).
    The last two words \textit{interim} and \textit{data} are therefore trimmed.
    There are $L$ layers in this BERT model.
    $Tok_i$ denotes the \textit{i-th} token of the tokenized news and $N$ denotes the maximum length of a headline.
    $T_{i, j}$ represents the hidden vector for the $j$-th transformer at layer $i$.
    The first [CLS] label is simply a special string we choose to signify the class of this headline. 
    This label is only used when fine-tuning the model, we leave it blank when we generate the embeddings.
    The original method to use BERT as classification model is to use
    directly [CLS] vector to represent the whole sentence, while we choose to
    use all hidden vectors at one layer to represent the sentence.
    The embedding generated with the model is shown in Equation \ref{eq:tuned_embedding}.}
  \label{fig:bert_tuned}
\end{figure}

Figure \ref{fig:bert_tuned} shows the structure of BERT model. It has $L$ layers and each layer has 
$N$ nodes, each node is a Transformer \citep{vaswani2017attention}. The first layer takes 
a tokenized headline as input and the BERT model generates $N \times L$ hidden vectors, 
denoted by $T_{i, j}$. The first token at the first layer is a special [CLS] token
reserved for fine-tuning.

For a specific downstream task, we can fine-tune the base BERT model suitable for this specific task. 
It means that we do not initialize the parameters of BERT model randomly, we use the pre-trained BERT model
as our initial state instead. We update the parameters in the base model with our domain-specific data.
This approach adds domain knowledge to the large scale language model \citep{yang2020finbert}, 
it can help the BERT model better understand the texts in specific situations.
In our case, we can fine-tune the base BERT model with our labelled financial news data
mentioned in Section \ref{subsec:data_labelling} to make it specialize in financial texts.

The fine-tuning process is straightforward. We input the class label (0 or 1) together with 
the tokenized headline into the first layer of a pre-trained BERT model. We set the target to 
the class label and loss function to cross-entropy, defined as:
\begin{equation}
  loss = \sum Y_{i}ln(P_{i}^{+}) + (1 - Y_{i})ln(1-P_{i}^{+})
\end{equation}
where $Y_{i}$ is the label for the news $i$ and $P_{i}^{+}$ denotes the probability
that the news $i$ is positive given by the model.
We use back-propagation \citep{hecht1992theory} to update the parameters 
in the model. We repeat such operation for several epochs until the loss converges.

In order to generate the contextualized embedding, we can either directly use the base BERT
model or use the fine-tuned model.
We first tokenize our headlines using SentencePiece tokenizer \citep{kudo2018sentencepiece}. 
If the number of tokens is smaller than $N$, 
we simply pad it to $N$ tokens by adding null tokens at the end.
If there are more than $N$ tokens, we remove the last tokens to make this sentence have 
exactly $N$ tokens.
We input these tokens into pre-trained BERT model as shown in 
Figure \ref{fig:bert_tuned} with the first token which stands for [CLS] label left blank.
Suppose that our BERT model has $L$ layers,
we can then generate $L$ different embeddings, denoted by $Emb_{base, l}$. We have,
\begin{equation}
  \label{eq:cross_entropy_loss}
    Emb_{base, l} = [T_{l, 2}, T_{l, 3}, ..., T_{l, N}]
\end{equation}
where $Emb_{base, l}$ is a $size(T_{l, i}) \times (N-1)$ matrix representing this headline
and $l$ denotes the layer from which we generate the embedding. We have $1 \leq l \leq L$.


Similarly, we can generate another $L$ embeddings from the fine-tuned BERT model,
denoted by $Emb_{tuned, l}$. We have,
\begin{equation}
  \label{eq:tuned_embedding}
    Emb_{tuned, l} = [T_{l, 2}^{'}, T_{l, 3}^{'}, ..., T_{l, N}^{'}]
\end{equation}
where $T_{l, i}^{'}$ denotes the hidden vector for the \textit{i-th}
token at the \textit{l-th} layer for the fine-tuned model.

\subsection{RNN Prediction Model}
\label{sec:rnn}

\begin{figure}[h]
  \centering
  \includegraphics[width=0.8\columnwidth]{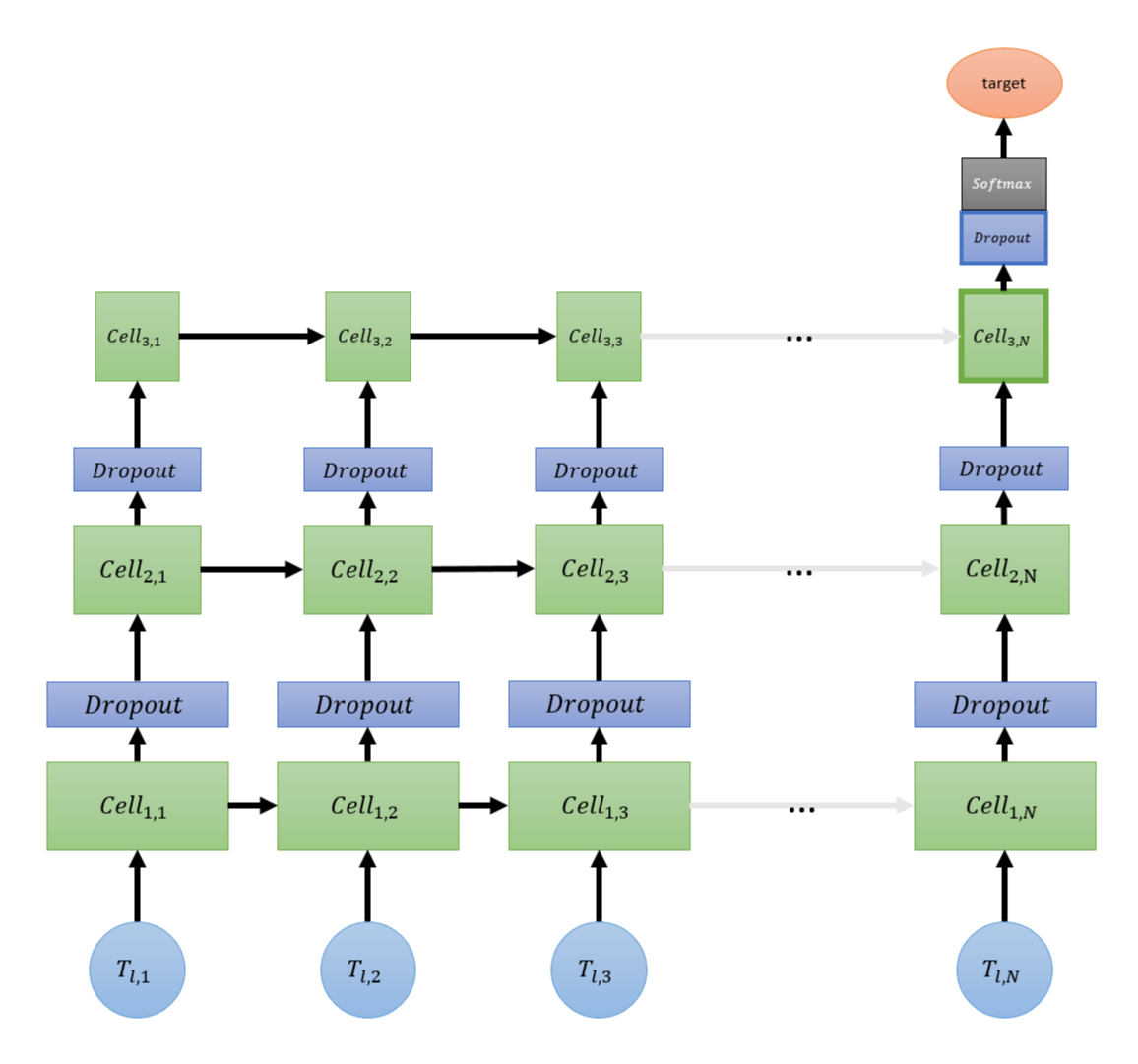}
  \caption{Structure of our RNN network. $Cell_{i, j}$ denotes the \textit{j-th} cell
    on the \textit{i-th} layer. The cell can be either Vanilla RNN, LSTM or GRU.
    In addition, the output size reduces when it approaches the last layer of the model.}
  \label{fig:rnn}
\end{figure}

The structure of our prediction model is simple and straightforward, it is shown in Figure \ref{fig:rnn}. 

There are several layers of cells which can either be Vanilla RNN \citep{cleeremans1989finite},
Long-Short Term Memory (LSTM) \citep{hochreiter1997long} or Gated Recurrent Unit (GRU) \citep{cho2014properties}. 
The size of output at each layer shrinks in order to reduce the dimension of our features gradually and
to make the remaining features more meaningful. Two neighbor layers are connected 
by a dropout to overcome the overfitting problem in the network.
At the end of the last layer, a softmax is added to calculate the probability for
each class based on the last vector on the last layer.

Suppose that we choose to use $Emb_{base, l}$ to represent a sentence. We first
initialize the parameters in the RNN model randomly and
input the embeddings of tokens ($T_{l,i}$) sequentially into the cells ($Cell_{1,i}$)
in the first layer of the recurrent neural network. At the same time, we set the target
to the corresponding label of the headline. We use the same cross-entropy loss mentioned in 
Equation \ref{eq:cross_entropy_loss} as our loss function. We use the same back propagation 
procedure in Section \ref{sec:ce} to update the parameters in the model until we have 
a stable loss.

\section{Experiments}
\label{sec:experiments}

In this section, we introduce our experiment setup and results in detail.
We also include the results of some baseline models to prove the effectiveness
of our model.
In addition to the final result, we add some ablation experiment results to
show the effect of some factors in our model.

\subsection{Training Setup}
\label{training_setup}

We use the pre-trained 12-layer, 768-dimension, 12-heads BERT 
model\footnote{This is the base pre-trained model published by Google, it is available at

https://storage.googleapis.com/bert\_models/2020\_02\_20/uncased\_L-12\_H-768\_A-12.zip}
to generate $E_{base, l}$, then we fine-tune this base model using our labelled dataset 
with the method mentioned in section 
\ref{subsec:data_labelling}.\footnote{We choose batch size: 32, 64, 128 and learning rate: 2e-6, 5e-6, 1e-5}
We choose the best-performing fine-tuned model to generate $E_{tuned, l}$.
Empirically, the layer of BERT used as embedding should not be too close
to the first layer, otherwise the contextualized embedding will be too
similar to the static embedding. Hence, we only test embeddings with
$l = L, L-1\ or\ L-2$. This choice will be discussed in detail in Section
\ref{sec:emblayer}.

The maximum length of a sentence is set to 32 tokens, as there are at most
29 tokens for all headlines in our dataset. If there are fewer tokens, 
we pad it to 32 with null tokens.

For our RNN model, the cells are set to be LSTM.
We use a four-layer single-directional 
RNN\footnote{The hidden size for each layer is set to 256, 128, 64, 32 respectively.} with a dropout rate of 50\%.

\subsection{Evaluation Metrics}
\label{subsec:evaluation_metrics}

Because of the huge volume of news that we receive daily, it is not realistic
for either human investors or systematic trading algorithms to react on all news.
Otherwise, we lose a considerable amount of transaction fees on the news which do not
significantly move the market. It is more logical that an investor first reads the news,
then buys or sells the stock if he thinks that the news can have substantial impact on the stock price.
If he thinks that the news is neutral, he will simply ignore the news.
In this type of neutral-insensitive scenario,
evaluating our model on all news is less meaningful. Instead, we evaluate our
model only on certain "extreme" news chosen based on their sentiment classification
results.

We define the score of a news, denoted by $S_{news}$:
\begin{equation}
  S_{news} = (P_{+}(news)-0.5) \times 2
\end{equation}
where $P_{+}(news)$ denotes the probability that this news belongs to
the positive class given by the prediction model. 
$S_{news}$ is therefore a value between -1 and 1.

We use $P_{n}$ to denote the $n^{th}$ percentile of all scores on the
\textbf{training} set. We can then choose the set on which we want to
evaluate our model. We define this set $E_{2n}$ by:
\begin{equation}
  \begin{aligned}
    E_{n}^{-} &= \{news | S_{news} < P_{n}\} \\
    E_{n}^{+} &= \{news | S_{news} > P_{100 - n}\} \\
    E_{2n} &= E_{n}^{-} \cup E_{n}^{+} \\
  \end{aligned}
\end{equation}

We assume that the distributions of scores on the training set and the test set are the same.
It should contain about $n\%$ highest-score news and $n\%$ lowest-score news from the test set.
We evaluate our model on these subsets of news instead of all news in the test set.

\bigskip
\noindent
\textbf{Standard Metrics}

Given a confusion matrix
$\bigl( \begin{smallmatrix}tp & fn\\ fp & tn\end{smallmatrix}\bigr)$
which contains the number of samples classified as true positive ($tp$), false positive ($fp$),
true negative ($tn$) and false negative ($fn$). We use both the accuracy and the Matthews 
Correlation Coefficient (MCC) \citep{matthews1975comparison} to evaluate our models. 
These two values are defined by:

\textbf{Accuracy}:
\begin{equation}
    \frac{tp + tn}{tp + tn + fp + fn}
\end{equation}

\textbf{Matthews Correlation Coefficient (MCC)}:
\begin{equation}
    \frac{tp \times tn - fp \times fn}{\sqrt{(tp + fp)(tp + fn)(tn + fp)(tn + fn)}}
\end{equation}

\bigskip
\noindent
\textbf{Trading Strategies}

However, those two metrics introduced above do not perfectly reflect
the reality, as the profits are quite different when the price
goes up significantly or mildly,
although they are both counted as true positive. Hence, it is
necessary to simulate these trades on real markets. We use two simple
trading strategies for simulations.

\bigskip
\textbf{Strategy 1 (S1)}.

We simply follow the strategy used by \citet{ke2019predicting}.

Before each market close, we search for all the news belonging to $E_{n}$ with a maximum age of $A$ days.
We group the selected news by stock and we calculate the average score for each stock.
We choose the 20 stocks with the highest scores, our target position for these stocks is \$T.
We also choose the 20 stocks with the lowest scores, our target position for these stocks is -\$T.

\bigskip
\textbf{Strategy 2 (S2)}.

The advantage of S1 is that it perfectly balances the long leg and the short leg.\footnote{
  The long leg means the total amount invested positively, the short leg means the total amount invested negatively}
As such, we have no exposure to the market, and it reduces the risk of the market movement.
However, the fallback of S1 is that it only focuses on the highest scores instead of considering all the stocks.
In this case, we will not be able to fully use our predictions.
Hence, we design the following strategy to solve this problem:

1. We first calculate the average score for each stock using the same method as described in S1.

2. We choose all the stocks with positive scores, $s_{i}^{+}$ denotes the score for the stock $i$.

3. The target position for the stock $i$ is $\$20T \times s_{i}^{+} / \sum_{j} s_{j}^{+}$.\footnote{
  The multiplier 20 is to guarantee the homogeneity with S1. We invest \$20T for each leg in both strategies.}

4. We invest in the stocks with negative scores in the same way. 
For a negatively scored stock $i$, we invest $-\$20T \times s_{i}^{-} / \sum_{j} s_{j}^{-}$.

This strategy not only uses all the available classification results, 
but also has no exposure to the market as S1, since the long position
and the short position are both $20T$.

\bigskip
To evaluate the performance of these strategies, we use the following two
commonly used indicators in finance.

\bigskip
\textbf{Annualized return}: defined by
\begin{equation}
    \frac{1}{N}\sum_{t=1}^{N}r_{t} \times D
\end{equation}

where $D$ is the number of trading days in one 
year\footnote{For the sake of simplicity, we choose 250 as the number of trading days for one year.},
and $r_{t}$ denotes the daily return of the portfolio for the day $t$, defined by 
the ratio of the profit on day $t$ to the total position on day $t$.

\bigskip
\textbf{Annualized Sharpe Ratio}: defined by
\begin{equation}
    \frac{\overline{\textbf{r}}}{\sigma(\textbf{r})} \times \sqrt{D}
\end{equation}

$\overline{\textbf{r}}$ denotes the mean of all the $r_{i}$ 
and $\sigma(\textbf{r})$ represents the standard deviation all the $r_{i}$.

\subsection{Baselines and Proposed Models}

We use the following models as baselines.

\begin{itemize}
  \item \textbf{NBC} \citep{maron1961automatic}: \textit{Naive Bayes Classifier}

        One of the most traditional language classification models based on word frequency.
  \item \textbf{SSESTM} \citep{ke2019predicting}:  \textit{Supervised Sentiment Extraction via Screening and Topic Modeling}.

  A regression model based on word frequency and stock returns.
  \item \textbf{Bloomberg} (Proprietary): \textit{Bloomberg Sentiment Score}
  
  The sentiment score from Bloomberg's proprietary model, 
  which comes along with the Bloomberg News dataset.
  An example of this sentiment score is shown in Table \ref{tab:data_sample}.
  \item \textbf{BERT} \citep{devlin2018bert}: \textit{Bidirectional Encoder Representations from Transformers}

  A general and powerful language model for a wide range of NLP tasks.
  We directly use the [CLS] label as the final prediction, as proposed by the author.
  \item \textbf{FinBERT} \citep{yang2020finbert}: \textit{Financial Sentiment Analysis with BERT}

  The same structure as the BERT model but pre-trained with financial domain-specific data.
\end{itemize}

To make a detailed analysis of the improvement brought by our proposed models, in addition to
the final version of our model (FT-CE-RNN), we add two other intermediate variants of our RNN model.

\begin{itemize}
  \item \textbf{RNN}: \textit{Recurrent Neural Network}

  The recurrent neural network introduced in Section \ref{sec:rnn}.
  Instead of using contextualized embeddings, we use the static Word2Vec embedding as its first layer.
  \item \textbf{CE-RNN}: \textit{Contextualized Embedding - Recurrent Neural Network}.

  The network structure is the same as RNN, but we use
  contextualized embedding generated from base BERT model instead of Word2Vec.
  \item \textbf{FT-CE-RNN}: \textit{Fine-Tuned - Contextualized Embedding - Recurrent Neural Network}

  The same RNN using contextualized embedding generated from fine-tuned BERT.
\end{itemize}

\subsection{Results}
\label{subsec:results}

\begin{figure}[h]
  \centering
  \subfigure[Accuracy]{\includegraphics[width=0.45\linewidth]{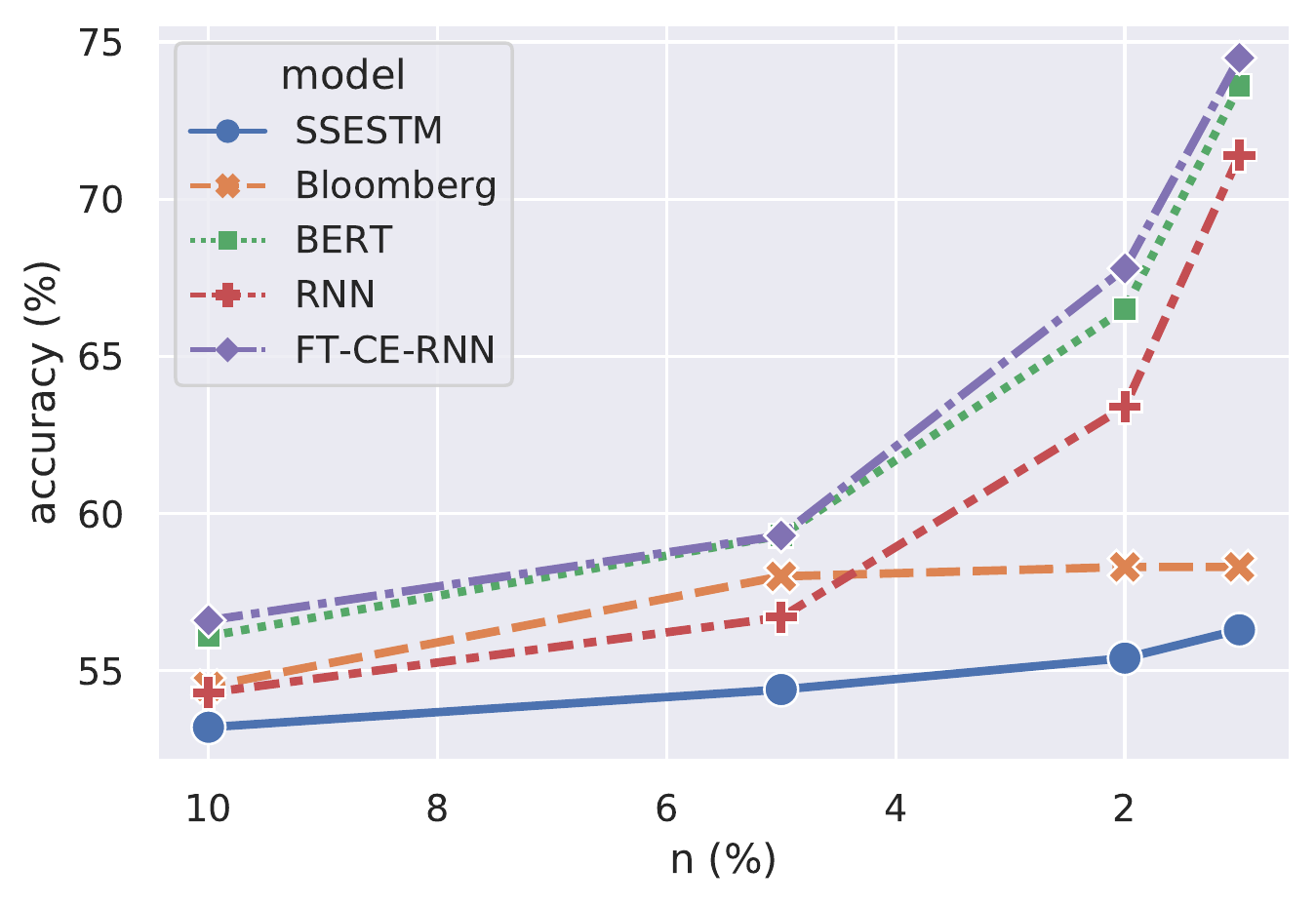}}
  \subfigure[MCC]{\includegraphics[width=0.45\linewidth]{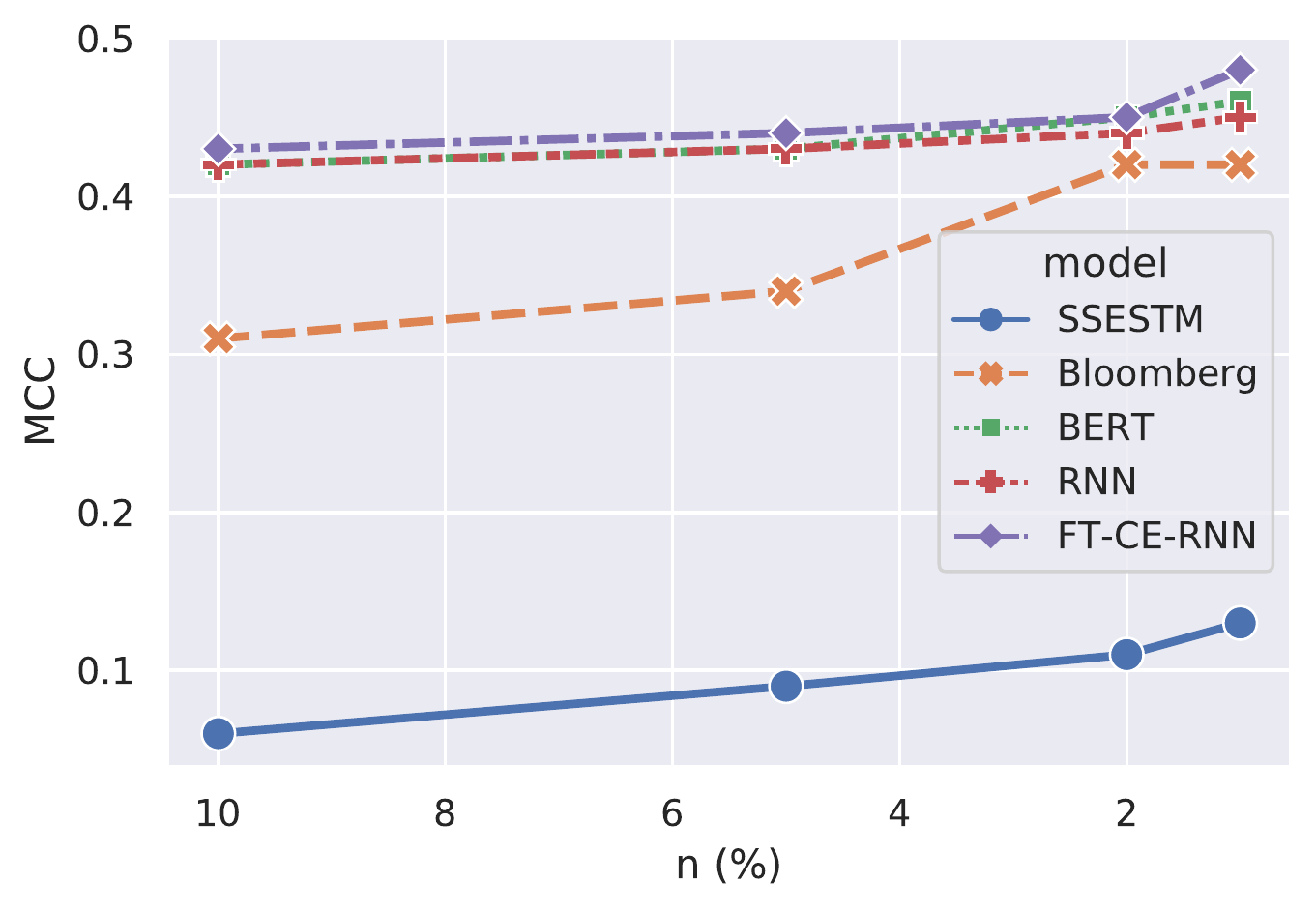}}
  \caption{Accuracy and MCC results of different models varied with $E_n$, the horizontal axis represents the value of $n$}%
  \label{fig:acc_mcc}
\end{figure}

The detailed results for standard metrics are shown in Table \ref{tab:res_acc} and Figure \ref{fig:acc_mcc}.
We also list the results from trading simulations in Table \ref{tab:res_strat}.

\begin{table*}[h]
  \begin{center}    
  \tbl{Performance of baseline models and our proposed RNN variants evaluated in accuracy and MCC}
  {
    \begin{tabular}{ccccccccc}
    \toprule
      & \multicolumn{2}{c}{1\%$^{\rm a}$} & \multicolumn{2}{c}{2\%} & \multicolumn{2}{c}{5\%} & \multicolumn{2}{c}{10\%} \\
    \midrule
      & Acc & MCC & Acc & MCC & Acc & MCC & Acc & MCC \\
    \midrule
    NBC & 59.8 & 0.2 & 56.1 & 0.12 & 54.3 & 0.09 & 53.4 & 0.07 \\
    SSESTM & 56.3 & 0.13 & 55.4 & 0.11 & 54.4 & 0.09 & 53.2 & 0.06 \\
    Bloomberg $^{\rm b}$ & 58.3 & 0.42 & 58.3 & 0.42 & 58.0 & 0.34 & 54.5 & 0.31 \\
    BERT & 73.6 & 0.46 & 66.5 & \textbf{0.45} & \textbf{59.3} & 0.43 & 56.1 & 0.42 \\
    FinBERT & 73.9 & 0.46 & 66.5 & \textbf{0.45} & 59.2 & 0.43 & 55.6 & 0.42 \\
    \midrule
    RNN & 71.4 & 0.45 & 63.4 & 0.44 & 56.7 & 0.43 & 54.3 & 0.42 \\
    CE-RNN & 70.9 & 0.42 & 63.8 & 0.28 & 57.9 & 0.16 & 54.7 & 0.09 \\
    FT-CE-RNN & \textbf{74.5} & \textbf{0.48} & \textbf{67.8} & \textbf{0.45} & \textbf{59.3} & \textbf{0.44} & \textbf{56.6} & \textbf{0.43} \\
    \bottomrule
    \end{tabular}}
    \label{tab:res_acc}%
  \end{center}
  \tabnote{$^{\rm a}$1\% signifies that we test our result on $E_{1}$, which includes about 1\% of all news on the test set.}
  \tabnote{$^{\rm b}$In Bloomberg dataset, we have about 4\% of the news with a maximum level score, therefore we have the same result for $E_{1}$ and $E_{2}$.}
\end{table*}%

\begin{table*}[h]
  \begin{center}    
  \tbl{Performance of baselines and proposed models in trading simulations$^{\rm a}$}
  {
    \begin{tabular}{cccccccc}
    \toprule
    \multirow{2}[4]{*}{Strategy} & \multirow{2}[4]{*}{Model} & \multicolumn{2}{c}{1\%} & \multicolumn{2}{c}{2\%} & \multicolumn{2}{c}{5\%} \\
\cmidrule{3-8}      &   & Ret.$^{\rm b}$ & Sharpe & Ret. & Sharpe & Ret. & Sharpe \\
    \midrule
    \multirow{8}[2]{*}{S1} & NBC & 9.61 & 1.09 & 2.35 & 0.26 & 2.44 & 0.27 \\
      & SSESTM & 2.39 & 0.41 & 1.57 & 0.24 & 2.74 & 0.35 \\
      & Bloomberg & 8.83 & 1.19 & 8.83 & 1.19 & 8.03 & 1.10 \\
      & BERT & 9.33 & 1.42 & 8.08 & 1.11 & 8.21 & 1.09 \\
      & FinBERT & 8.83 & 1.22 & 8.29 & 1.10 & 7.86 & 1.13 \\
      \cmidrule{2-8} & RNN & 9.86 & 1.43 & 8.06 & 1.13 & 6.43 & 0.89 \\
      & CE-RNN & 8.39 & 1.07 & 7.31 & 0.99 & 7.62 & 1.01 \\
      & FT-CE-RNN & 10.75 & 1.50 & 12.31 & 1.70 & 11.32 & 1.50 \\
    \midrule
    \multirow{8}[2]{*}{S2} & NBC & 7.93 & 0.62 & 0.57 & 0.06 & 0.80 & 0.15 \\
      & SSESTM & 4.75 & 0.47 & 3.02 & 0.35 & 2.76 & 0.38 \\
      & Bloomberg & 10.76 & 1.61 & 10.76 & 1.61 & 9.82 & 1.56 \\
      & BERT & 13.10 & 1.42 & 11.42 & 1.56 & 9.87 & 1.53 \\
      & FinBERT & 11.35 & 1.24 & 9.85 & 1.21 & 12.85 & 1.97 \\
      \cmidrule{2-8} & RNN & 17.53 & 1.75 & 15.47 & 1.72 & 12.70 & 1.79 \\
      & CE-RNN & 12.58 & 1.33 & 10.37 & 1.25 & 8.29 & 1.05 \\
      & FT-CE-RNN & \textbf{19.72} & \textbf{2.11} & \textbf{18.39} & \textbf{2.49} & \textbf{15.01} & \textbf{2.31} \\
    \bottomrule
  \end{tabular}}
  \label{tab:res_strat}%
  \end{center}
  \tabnote{$^{\rm a}$The simulation does not consider transaction costs.}
    \tabnote{$^{\rm b}$The annualized return, presented in percent.}
\end{table*}%

We find that in terms of accuracy, our FT-CE-RNN outperforms all the other baselines models.
Especially, comparing the accuracy of RNN and FT-CE-RNN,
there is an improvement of 4.1\% when we test on the 1\% most extreme news ($E_{1}$). This result shows
the power of contextualized embedding against static embedding. We also notice that there is
an improvement of 0.9\% compared with BERT result. This result explains that using all
hidden vectors instead of only one [CLS] vector helps improve the result. However, if we 
directly use the embedding from the base BERT (CE-RNN) instead of the fined-tuned BERT (FT-CE-RNN),
there is a clear disadvantage. This result shows the necessity of including the domain knowledge in 
the embeddings.

In addition, we observe that all our models have a significant margin compared with the Bloomberg and SSESTM
sentiment score, which is completely independent of our data labelling and modelling process, this result 
can prove the efficiency of our models compared with other widely used 
models in the industry.

In our trading simulations, we use a look-back window of 5 days ($A=5 \ days$).
We find that the result of our FT-CE-RNN outperforms all other
models in most of the cases. The most significant improvement is on Sharpe Ratio. When trading
on the 1\% most extreme news using strategy S2, there is an improvement of 0.69 (49\%) in Sharpe ratio 
and an improvement of 6.62 (51\%) in return if we compare BERT and FT-CE-RNN. 
It means that our model using contextualized embedding is not only more profitable but also more stable.

We can also find that S2 performs better than S1, since S2 uses all the signals while S1 only 
uses the signals on the top/bottom stocks. This proves that our classification is valid for 
most of the stocks, making it a robust method.

\begin{figure}[h]
  \centering
  \includegraphics[width=0.8\columnwidth]{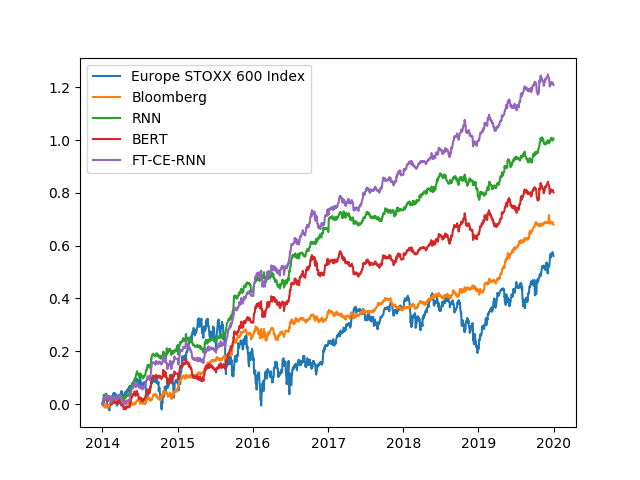}
  \caption{Trading simulation result in absolute profit.
    This trading simulation is run on 1\% most extreme news ($E_{1}$) using strategy S2.
    We also include the market return, represented by Europe STOXX 600 Index. 
    We can see our models largely exceed the market return and the Bloomberg sentiment score.}
  \label{fig:backtest}
\end{figure}

An example of trading simulation is shown in Figure \ref{fig:backtest}. It shows how our profit
evolves with time. We observe that FT-CE-RNN is not only better on profitability and stability
but also on the absolute profit in dollars.

\subsection{Transaction costs}
\begin{table}[h]
  \begin{center}
  \tbl{Trading simulation results with transaction costs$^{\rm a}$}
  {
    \begin{tabular}{ccccc}
    \toprule
    \multirow{2}[4]{*}{Model} & \multicolumn{2}{c}{No cost} & \multicolumn{2}{c}{With cost} \\
\cmidrule{2-5}      & Ret. & Sharpe & Ret. & Sharpe \\
    \midrule
    NBC & 7.93 & 0.62 & 2.09 & 0.16 \\
    SSESTM & 4.75 & 0.47 & -1.80 & -0.18 \\
    Bloomberg & 10.76 & 1.61 & 4.49 & 0.67 \\
    BERT & 13.10 & 1.42 & 5.98 & 0.65 \\
    FinBERT & 11.35 & 1.24 & 3.88 & 0.42 \\
    \midrule
    RNN & 17.53 & 1.75 & 11.57 & 1.15 \\
    CE-RNN & 12.58 & 1.33 & 6.15 & 0.65 \\
    FT-CE-RNN & \textbf{19.72} & \textbf{2.11} & \textbf{12.60} & \textbf{1.35} \\
    \bottomrule
    \end{tabular}}
    \label{tab:res_strat_costs}%
  \end{center}
    \tabnote{$^{\rm a}$The result is obtained based on $E_{1}$ test set using the trading strategy S2.}
\end{table}%

Our trading simulations ignore transaction costs thus far, since the primary goal of this 
research is to prove the effectiveness of our sentiment model with contextualized embedding. 
The transaction costs have no impact 
on the result because all the models are in the same no-cost environment.

That said, applying this model in real-life trading is another separate but interesting question.
To understand the real gain of our FT-CE-RNN model for the asset management, we rerun our trading 
simulations with transaction costs.

In our simulations, we assume a transaction cost of 4bps\footnote{basis points, $10^{-4}$} 
proportional to the daily turnover\footnote{defined as $|pos_{i} - pos_{i-1}|$ for day $i$}. 
The simulation results with transaction costs are shown in Table \ref{tab:res_strat_costs}.

Although the transaction costs cut our profits significantly, we can still have a profitable
margin when using our FT-CE-RNN model.

\subsection{Effect of Embedding Layer}
\label{sec:emblayer}

In this section, we discuss our choice of BERT hidden layer to be used as embedding.

Empirically, the final layer of BERT model should be used to generate our contextualized
embedding as the final layer is more "mature" and contains more information compared
with other layers which are closer to the first layer. However, our result
listed in Table \ref{tab:emb_layer} shows the opposite.

\begin{table}[h]
  \begin{center}
    \tbl{Accuracy result using different layers of BERT model as contextualized embedding$^{\rm a}$}
    {
  \begin{tabular}{cccc}
    \toprule
    Embedding layer & $L$$^{\rm b}$           & $L-1$         & $L-2$ \\
    \midrule
    CE-RNN       & \textbf{70.9}  & 68.1 & 63.3  \\
    FT-CE-RNN    & 73.0 & \textbf{74.5}  & 66.7  \\
    \bottomrule
  \end{tabular}}
  \label{tab:emb_layer}%
\end{center}
  \tabnote{$^{\rm a}$The result is acquired on the $E_{1}$ test set.}
  \tabnote{$^{\rm b}$$L$ denotes the last layer of the BERT model, $L-n$ represents the $n$-th last layer of the BERT model.}
\end{table}%

We find that the best result for CE-RNN is acquired when we use layer $L$ while the best result
for FT-CE-RNN is obtained when using layer $L - 1$. The reason for this phenomenon is that 
the last layer of the fine-tuned BERT is biased towards the classification result, since 
the goal for the fine-tuning process is to make the first token of the last layer close to the classification target. 
If we use the last layer of the fine-tuned BERT as the input for RNN, we are simply replicating the classification process of the 
BERT, instead of improving the result. Using one deeper layer ($L-1$) helps reduce this bias \citep{xiao2018bertservice}.
However, the base BERT does not have this bias on the last layer since it has no previous knowledge on 
the training set. This explains why CE-RNN has a better performance when using the embedding from the last layer ($L$).

If we use an even lower layer to generate contextualized embedding, such as $L-2$, the performance
declines as it is too close to the embedding layer and lacks contextualized characteristics.

\subsection{Effect of Classification Classes}
\label{subsec:multi_class_result}

During our initial researches, we also explored the possibility of using a 3-class classification instead of a 2-class classification.
It means that we do not classify a news into a positive or a negative news, we classify if a news is either positive, 
negative or neutral. This is the method adopted in the Bloomberg's proprietary model.

\begin{table}[h]
  \begin{center}
  \tbl{The accuracy result for a 2-class classification model and a 3-class classification model$^{\rm a}$$^{\rm b}$}
  {
  \begin{tabular}{ccccc}
    \toprule
    Acc & 1\% & 2\% & 5\% & 10\% \\
    \midrule
    2-class & 74.5 & 67.8 & 59.3 & 56.6 \\
    3-class & 61.0 & 58.4 & 55.5 & 53.6 \\
    \bottomrule
  \end{tabular}}
  \label{tab:3class}%
\end{center}
  \tabnote{$^{\rm a}$The result is based on the test set $E_{1}$.}
  \tabnote{$^{\rm b}$For 3-class classification, we choose the $n/2$\% largest scores
  for the positive class and the $n/2$\% largest scores for the negative class as our $E_{n}$.
  The guarantees the same number of the news considered in both cases.}
\end{table}%

The result of using a 3-class classification model is shown in Table \ref{tab:3class}. We find a significant 
worse performance if we add another possibility to our model. This is because a 3-class model supposes 
a clear difference between the market-moving news and the neutral news, however, this is not always 
the case. It is not obvious to find a threshold, above which the news is positive and below 
which the news is neutral. In this scenario, we are not able to construct a clear training set
for our model to learn the difference between a neutral news and a market-moving news.

Hence, in our final model, we decide to classify the news into two classes instead of three.

\subsection{Qualitative Analysis of the Classification Result}

We analyze the news in $E_{1}$ to see if there is any pattern, for example,
some frequent words in them.
We include the 50 most frequent words appeared in $E^{+}_{0.5}$ and 50 most 
frequent words in $E^{-}_{0.5}$ in Appendix \ref{sec:frequent_words}, along
with their frequencies.
We exclude all the stopwords in English, such as \textit{to}, \textit{for}, \textit{a}, etc.

We can find that for the news identified as the most positive, some common words include
\textit{buy}, \textit{upgrade}, \textit{raise}, etc. For the news identified as the
most negative, \textit{downgrade}, \textit{cut} and \textit{miss} are among common words.
These are also logical keywords for the humans,
making the result from the neural network intuitive.

We can also find in this collection that there are also some less natural words, such as
\textit{fly}, \textit{say} \textit{neutral}, etc. 
However, as these words appear in both categories, the effect of such words is neutralized
if we empirically assume the effect of a word is its positive impact minus its negative impact.

This result is similar to the result we get from the word frequency-based method, 
such as NBC and SSESTM. However, we demonstrate that our FT-CE-RNN model is significantly
more powerful than these two baseline models (Section \ref{subsec:results}).
This phenomenon implies that our model is capable of capturing complex
information in the news on top of the word frequency.

\section{Conclusion}

We build the whole pipeline for the stock movement prediction task with headlines from financial news,
including labeling the news, generating contextualized embedding, training a neural network model,
validating the model with various metrics and
building trading strategies based on the model output.

We design a FT-CE-RNN model which uses fine-tuned contextualized embeddings from BERT 
instead of the traditional static embeddings. 
We also introduce our new evaluation metrics focusing on market-moving news, 
which are more suitable for asset manager's needs.

Through various experiments on the Bloomberg News dataset, 
we demonstrate the effectiveness of our FT-CE-RNN model.
We find a better performance, in both accuracy and trading simulations, 
than other widely used baseline models.
We also include other ablation studies to discuss the choice of 
some important parameters and to demonstrate the intuitiveness of the result.

In the future, we will continue our research on the stock movement prediction using 
natural language processing methods based on longer texts (such as earning call transcripts,
financial reports, etc.) instead of the headlines. 
By using more information, we aim to build a model 
which helps achieve better stock movement prediction result.

\section*{Acknowledgments}
The authors gratefully acknowledge the financial support of the Chaire 
\textit{Machine Learning \& Systematic Methods}
and the Chaire \textit{Analytics and Models for Regulation}.

\newpage
\bibliographystyle{rQUF}
\bibliography{rQUFguide}

\begin{thebibliography}{57}
\providecommand{\natexlab}[1]{#1}
\providecommand{\noopsort}[1]{}
\providecommand{\printfirst}[2]{#1}
\providecommand{\singleletter}[1]{#1}
\providecommand{\switchargs}[2]{#2#1}

\bibitem[\protect\citeauthoryear{Adebiyi
  {\itshape{et~al.}}}{2012}]{adebiyi2012stock}
Adebiyi, A.A., Ayo, C.K., Adebiyi, M.O. and Otokiti, S.O., Stock price
  prediction using neural network with hybridized market indicators. {\itshape
  Journal of Emerging Trends in Computing and Information Sciences}, 2012,
  \textbf{3}, 1--9.

\bibitem[\protect\citeauthoryear{Ariyo
  {\itshape{et~al.}}}{2014}]{ariyo2014stock}
Ariyo, A.A., Adewumi, A.O. and Ayo, C.K., Stock price prediction using the
  ARIMA model. In {\itshape Proceedings of the }{\itshape 2014 UKSim-AMSS 16th
  International Conference on Computer Modelling and Simulation}, pp. 106--112,
  2014.

\bibitem[\protect\citeauthoryear{Bai and Pukthuanthong}{2020}]{bai2020machine}
Bai, Y. and Pukthuanthong, K., Machine Learning Classification Methods and
  Portfolio Allocation: An Examination of Market Efficiency. {\itshape
  Available at SSRN 3665051}, 2020.

\bibitem[\protect\citeauthoryear{Bhandari}{1988}]{bhandari1988debt}
Bhandari, L.C., Debt/equity ratio and expected common stock returns: Empirical
  evidence. {\itshape The journal of finance}, 1988, \textbf{43}, 507--528.

\bibitem[\protect\citeauthoryear{Cer
  {\itshape{et~al.}}}{2018}]{cer2018universal}
Cer, D., Yang, Y., Kong, S.y., Hua, N., Limtiaco, N., John, R.S., Constant, N.,
  Guajardo-C{\'e}spedes, M., Yuan, S., Tar, C. {\itshape et~al.}, Universal
  sentence encoder. {\itshape arXiv preprint arXiv:1803.11175}, 2018.

\bibitem[\protect\citeauthoryear{Chen and Ge}{2019}]{chen2019exploring}
Chen, S. and Ge, L., Exploring the attention mechanism in LSTM-based Hong Kong
  stock price movement prediction. {\itshape Quantitative Finance}, 2019,
  \textbf{19}, 1507--1515.

\bibitem[\protect\citeauthoryear{Cho
  {\itshape{et~al.}}}{2014}]{cho2014properties}
Cho, K., Van~Merri{\"e}nboer, B., Bahdanau, D. and Bengio, Y., On the
  properties of neural machine translation: Encoder-decoder approaches.
  {\itshape arXiv preprint arXiv:1409.1259}, 2014.

\bibitem[\protect\citeauthoryear{Cleeremans
  {\itshape{et~al.}}}{1989}]{cleeremans1989finite}
Cleeremans, A., Servan-Schreiber, D. and McClelland, J.L., Finite state
  automata and simple recurrent networks. {\itshape Neural computation}, 1989,
  \textbf{1}, 372--381.

\bibitem[\protect\citeauthoryear{Cooper
  {\itshape{et~al.}}}{2016}]{cooper2016performance}
Cooper, M.J., Gulen, H. and Rau, P.R., Performance for pay? The relation
  between CEO incentive compensation and future stock price performance.
  {\itshape The Relation Between CEO Incentive Compensation and Future Stock
  Price Performance (November 1, 2016)}, 2016.

\bibitem[\protect\citeauthoryear{Coqueret}{2020}]{coqueret2020stock}
Coqueret, G., Stock-specific sentiment and return predictability. {\itshape
  Quantitative Finance}, 2020, \textbf{20}, 1531--1551.

\bibitem[\protect\citeauthoryear{Del~Corro and
  Hoffart}{2020}]{del2020unsupervised}
Del~Corro, L. and Hoffart, J., Unsupervised Extraction of Market Moving Events
  with Neural Attention. {\itshape arXiv preprint arXiv:2001.09466}, 2020.

\bibitem[\protect\citeauthoryear{Devlin
  {\itshape{et~al.}}}{2018}]{devlin2018bert}
Devlin, J., Chang, M., Lee, K. and Toutanova, K., {BERT:} Pre-training of Deep
  Bidirectional Transformers for Language Understanding. {\itshape CoRR}, 2018,
  \textbf{abs/1810.04805}.

\bibitem[\protect\citeauthoryear{Ding {\itshape{et~al.}}}{2014}]{ding2014using}
Ding, X., Zhang, Y., Liu, T. and Duan, J., Using structured events to predict
  stock price movement: An empirical investigation. In {\itshape Proceedings of
  the }{\itshape Proceedings of the 2014 Conference on Empirical Methods in
  Natural Language Processing (EMNLP)}, pp. 1415--1425, 2014.

\bibitem[\protect\citeauthoryear{Ding {\itshape{et~al.}}}{2015}]{ding2015deep}
Ding, X., Zhang, Y., Liu, T. and Duan, J., Deep learning for event-driven stock
  prediction. In {\itshape Proceedings of the }{\itshape Twenty-Fourth
  International Joint Conference on Artificial Intelligence}, 2015.

\bibitem[\protect\citeauthoryear{Donaldson and
  Storeygard}{2016}]{donaldson2016view}
Donaldson, D. and Storeygard, A., The view from above: Applications of
  satellite data in economics. {\itshape Journal of Economic Perspectives},
  2016, \textbf{30}, 171--98.

\bibitem[\protect\citeauthoryear{Fama}{1965}]{fama1965behavior}
Fama, E.F., The behavior of stock-market prices. {\itshape The journal of
  Business}, 1965, \textbf{38}, 34--105.

\bibitem[\protect\citeauthoryear{Fedyk}{2018}]{fedyk2018front}
Fedyk, A., Front page news: The effect of news positioning on financial
  markets. Technical report, working paper, 2018.

\bibitem[\protect\citeauthoryear{Hecht-Nielsen}{1992}]{hecht1992theory}
Hecht-Nielsen, R., Theory of the backpropagation neural network. In {\itshape
  Neural networks for perception}, pp. 65--93, 1992, Elsevier.

\bibitem[\protect\citeauthoryear{Hochreiter and
  Schmidhuber}{1997}]{hochreiter1997long}
Hochreiter, S. and Schmidhuber, J., Long short-term memory. {\itshape Neural
  computation}, 1997, \textbf{9}, 1735--1780.

\bibitem[\protect\citeauthoryear{Hu {\itshape{et~al.}}}{2018}]{hu2018listening}
Hu, Z., Liu, W., Bian, J., Liu, X. and Liu, T.Y., Listening to chaotic
  whispers: A deep learning framework for news-oriented stock trend prediction.
  In {\itshape Proceedings of the }{\itshape Proceedings of the Eleventh ACM
  International Conference on Web Search and Data Mining}, pp. 261--269, 2018.

\bibitem[\protect\citeauthoryear{Jiang
  {\itshape{et~al.}}}{2018}]{jiang2018short}
Jiang, Z.Q., Wang, G.J., Canabarro, A., Podobnik, B., Xie, C., Stanley, H.E.
  and Zhou, W.X., Short term prediction of extreme returns based on the
  recurrence interval analysis. {\itshape Quantitative Finance}, 2018,
  \textbf{18}, 353--370.

\bibitem[\protect\citeauthoryear{Jones}{1972}]{jones1972statistical}
Jones, K.S., A statistical interpretation of term specificity and its
  application in retrieval. {\itshape Journal of documentation}, 1972.

\bibitem[\protect\citeauthoryear{Ke
  {\itshape{et~al.}}}{2019}]{ke2019predicting}
Ke, Z.T., Kelly, B.T. and Xiu, D., Predicting returns with text data. Technical
  report, National Bureau of Economic Research, 2019.

\bibitem[\protect\citeauthoryear{Kohara
  {\itshape{et~al.}}}{1997}]{kohara1997stock}
Kohara, K., Ishikawa, T., Fukuhara, Y. and Nakamura, Y., Stock price prediction
  using prior knowledge and neural networks. {\itshape Intelligent Systems in
  Accounting, Finance \& Management}, 1997, \textbf{6}, 11--22.

\bibitem[\protect\citeauthoryear{Kraft and Kraft}{1977}]{kraft1977determinants}
Kraft, J. and Kraft, A., Determinants of common stock prices: a time series
  analysis. {\itshape The journal of finance}, 1977, \textbf{32}, 417--425.

\bibitem[\protect\citeauthoryear{Kroujiline
  {\itshape{et~al.}}}{2016}]{kroujiline2016forecasting}
Kroujiline, D., Gusev, M., Ushanov, D., Sharov, S.V. and Govorkov, B.,
  Forecasting stock market returns over multiple time horizons. {\itshape
  Quantitative Finance}, 2016, \textbf{16}, 1695--1712.

\bibitem[\protect\citeauthoryear{Kudo and
  Richardson}{2018}]{kudo2018sentencepiece}
Kudo, T. and Richardson, J., SentencePiece: {A} simple and language independent
  subword tokenizer and detokenizer for Neural Text Processing. {\itshape
  CoRR}, 2018, \textbf{abs/1808.06226}.

\bibitem[\protect\citeauthoryear{Li
  {\itshape{et~al.}}}{2020}]{li2020identifying}
Li, J., Li, G., Zhu, X. and Yao, Y., Identifying the influential factors of
  commodity futures prices through a new text mining approach. {\itshape
  Quantitative Finance}, 2020, \textbf{20}, 1967--1981.

\bibitem[\protect\citeauthoryear{Liu {\itshape{et~al.}}}{2007}]{liu2007cash}
Liu, J., Nissim, D. and Thomas, J., Is cash flow king in valuations?. {\itshape
  Financial Analysts Journal}, 2007, \textbf{63}, 56--68.

\bibitem[\protect\citeauthoryear{Luss and
  d'Aspremont}{2015}]{luss2015predicting}
Luss, R. and d'Aspremont, A., Predicting abnormal returns from news using text
  classification. {\itshape Quantitative Finance}, 2015, \textbf{15},
  999--1012.

\bibitem[\protect\citeauthoryear{M{\"a}kinen
  {\itshape{et~al.}}}{2019}]{makinen2019forecasting}
M{\"a}kinen, Y., Kanniainen, J., Gabbouj, M. and Iosifidis, A., Forecasting
  jump arrivals in stock prices: new attention-based network architecture using
  limit order book data. {\itshape Quantitative Finance}, 2019, \textbf{19},
  2033--2050.

\bibitem[\protect\citeauthoryear{Maron}{1961}]{maron1961automatic}
Maron, M.E., Automatic indexing: an experimental inquiry. {\itshape Journal of
  the ACM (JACM)}, 1961, \textbf{8}, 404--417.

\bibitem[\protect\citeauthoryear{Matthews}{1975}]{matthews1975comparison}
Matthews, B.W., Comparison of the predicted and observed secondary structure of
  T4 phage lysozyme. {\itshape Biochimica et Biophysica Acta (BBA)-Protein
  Structure}, 1975, \textbf{405}, 442--451.

\bibitem[\protect\citeauthoryear{Mikolov
  {\itshape{et~al.}}}{2013}]{mikolov2013distributed}
Mikolov, T., Sutskever, I., Chen, K., Corrado, G.S. and Dean, J., Distributed
  representations of words and phrases and their compositionality. In {\itshape
  Proceedings of the }{\itshape Advances in neural information processing
  systems}, pp. 3111--3119, 2013.

\bibitem[\protect\citeauthoryear{Nguyen and Shirai}{2015}]{nguyen2015topic}
Nguyen, T.H. and Shirai, K., Topic modeling based sentiment analysis on social
  media for stock market prediction. In {\itshape Proceedings of the }{\itshape
  Proceedings of the 53rd Annual Meeting of the Association for Computational
  Linguistics and the 7th International Joint Conference on Natural Language
  Processing (Volume 1: Long Papers)}, Vol. ~1, pp. 1354--1364, 2015.

\bibitem[\protect\citeauthoryear{Nonejad}{2021}]{nonejad2021bayesian}
Nonejad, N., Bayesian model averaging and the conditional volatility process:
  an application to predicting aggregate equity returns by conditioning on
  economic variables. {\itshape Quantitative Finance}, 2021, pp. 1--25.

\bibitem[\protect\citeauthoryear{Obaid and
  Pukthuanthong}{2021}]{obaid2021picture}
Obaid, K. and Pukthuanthong, K., A picture is worth a thousand words: Measuring
  investor sentiment by combining machine learning and photos from news.
  {\itshape Journal of Financial Economics}, 2021.

\bibitem[\protect\citeauthoryear{Oliveira
  {\itshape{et~al.}}}{2013}]{oliveira2013some}
Oliveira, N., Cortez, P. and Areal, N., Some experiments on modeling stock
  market behavior using investor sentiment analysis and posting volume from
  Twitter. In {\itshape Proceedings of the }{\itshape Proceedings of the 3rd
  International Conference on Web Intelligence, Mining and Semantics}, p.~31,
  2013.

\bibitem[\protect\citeauthoryear{Pagolu
  {\itshape{et~al.}}}{2016}]{pagolu2016sentiment}
Pagolu, V.S., Reddy, K.N., Panda, G. and Majhi, B., Sentiment analysis of
  Twitter data for predicting stock market movements. In {\itshape Proceedings
  of the }{\itshape 2016 international conference on signal processing,
  communication, power and embedded system (SCOPES)}, pp. 1345--1350, 2016.

\bibitem[\protect\citeauthoryear{Patell}{1976}]{patell1976corporate}
Patell, J.M., Corporate forecasts of earnings per share and stock price
  behavior: Empirical test. {\itshape Journal of accounting research}, 1976,
  pp. 246--276.

\bibitem[\protect\citeauthoryear{Pennington
  {\itshape{et~al.}}}{2014}]{pennington2014glove}
Pennington, J., Socher, R. and Manning, C.D., GloVe: Global Vectors for Word
  Representation. In {\itshape Proceedings of the }{\itshape Empirical Methods
  in Natural Language Processing (EMNLP)}, pp. 1532--1543, 2014.

\bibitem[\protect\citeauthoryear{Peters
  {\itshape{et~al.}}}{2018}]{peters2018deep}
Peters, M.E., Neumann, M., Iyyer, M., Gardner, M., Clark, C., Lee, K. and
  Zettlemoyer, L., Deep contextualized word representations. {\itshape arXiv
  preprint arXiv:1802.05365}, 2018.

\bibitem[\protect\citeauthoryear{Rekabsaz
  {\itshape{et~al.}}}{2017}]{rekabsaz2017volatility}
Rekabsaz, N., Lupu, M., Baklanov, A., Hanbury, A., D{\"u}r, A. and Anderson,
  L., Volatility prediction using financial disclosures sentiments with word
  embedding-based ir models. {\itshape arXiv preprint arXiv:1702.01978}, 2017.

\bibitem[\protect\citeauthoryear{Schumaker and
  Chen}{2009}]{schumaker2009textual}
Schumaker, R.P. and Chen, H., Textual analysis of stock market prediction using
  breaking financial news: The AZFin text system. {\itshape ACM Transactions on
  Information Systems (TOIS)}, 2009, \textbf{27}, 12.

\bibitem[\protect\citeauthoryear{Si
  {\itshape{et~al.}}}{2013}]{si2013exploiting}
Si, J., Mukherjee, A., Liu, B., Li, Q., Li, H. and Deng, X., Exploiting topic
  based twitter sentiment for stock prediction. In {\itshape Proceedings of the
  }{\itshape Proceedings of the 51st Annual Meeting of the Association for
  Computational Linguistics (Volume 2: Short Papers)}, Vol. ~2, pp. 24--29,
  2013.

\bibitem[\protect\citeauthoryear{Sonsino and Shavit}{2014}]{sonsino2014return}
Sonsino, D. and Shavit, T., Return prediction and stock selection from
  unidentified historical data. {\itshape Quantitative Finance}, 2014,
  \textbf{14}, 641--655.

\bibitem[\protect\citeauthoryear{Stevens
  {\itshape{et~al.}}}{1946}]{stevens1946theory}
Stevens, S.S. {\itshape et~al.}, On the theory of scales of measurement. ,
  1946.

\bibitem[\protect\citeauthoryear{Tashiro
  {\itshape{et~al.}}}{2019}]{tashiro2019encoding}
Tashiro, D., Matsushima, H., Izumi, K. and Sakaji, H., Encoding of
  high-frequency order information and prediction of short-term stock price by
  deep learning. {\itshape Quantitative Finance}, 2019, \textbf{19},
  1499--1506.

\bibitem[\protect\citeauthoryear{Vaswani
  {\itshape{et~al.}}}{2017}]{vaswani2017attention}
Vaswani, A., Shazeer, N., Parmar, N., Uszkoreit, J., Jones, L., Gomez, A.N.,
  Kaiser, L. and Polosukhin, I., Attention is all you need. {\itshape arXiv
  preprint arXiv:1706.03762}, 2017.

\bibitem[\protect\citeauthoryear{Wan
  {\itshape{et~al.}}}{2021}]{wan2021sentiment}
Wan, X., Yang, J., Marinov, S., Calliess, J.P., Zohren, S. and Dong, X.,
  Sentiment correlation in financial news networks and associated market
  movements. {\itshape Scientific reports}, 2021, \textbf{11}, 1--12.

\bibitem[\protect\citeauthoryear{Wang {\itshape{et~al.}}}{2018}]{wang2018glue}
Wang, A., Singh, A., Michael, J., Hill, F., Levy, O. and Bowman, S.R., GLUE: A
  multi-task benchmark and analysis platform for natural language
  understanding. {\itshape arXiv preprint arXiv:1804.07461}, 2018.

\bibitem[\protect\citeauthoryear{Xiao}{2018}]{xiao2018bertservice}
Xiao, H., bert-as-service. \url{https://github.com/hanxiao/bert-as-service},
  2018.

\bibitem[\protect\citeauthoryear{Xie
  {\itshape{et~al.}}}{2013}]{xie2013semantic}
Xie, B., Passonneau, R., Wu, L. and Creamer, G.G., Semantic frames to predict
  stock price movement. In {\itshape Proceedings of the }{\itshape Proceedings
  of the 51st annual meeting of the association for computational linguistics},
  pp. 873--883, 2013.

\bibitem[\protect\citeauthoryear{Xu and Cohen}{2018}]{xu2018stock}
Xu, Y. and Cohen, S.B., Stock movement prediction from tweets and historical
  prices. In {\itshape Proceedings of the }{\itshape Proceedings of the 56th
  Annual Meeting of the Association for Computational Linguistics (Volume 1:
  Long Papers)}, pp. 1970--1979, 2018.

\bibitem[\protect\citeauthoryear{Yang
  {\itshape{et~al.}}}{2020}]{yang2020finbert}
Yang, Y., UY, M.C.S. and Huang, A., FinBERT: A Pretrained Language Model for
  Financial Communications. , 2020.

\bibitem[\protect\citeauthoryear{Yang {\itshape{et~al.}}}{2019}]{yang2019xlnet}
Yang, Z., Dai, Z., Yang, Y., Carbonell, J., Salakhutdinov, R. and Le, Q.V.,
  Xlnet: Generalized autoregressive pretraining for language understanding.
  {\itshape arXiv preprint arXiv:1906.08237}, 2019.

\bibitem[\protect\citeauthoryear{Zhang and Yan}{2018}]{zhang2018modelling}
Zhang, H. and Yan, C., Modelling fundamental analysis in portfolio selection.
  {\itshape Quantitative Finance}, 2018, \textbf{18}, 1315--1326.

\end{thebibliography}

\newpage
\appendices
\section{Frequent Words in Market Moving News}
\label{sec:frequent_words}

\begin{table}[htbp]
  \begin{center}
  \tbl{The most frequent words in the most positvely scored news 
  and the most negatively scored news}{
    \begin{tabular}{cc|cc}
      \toprule
 \multicolumn{2}{c|}{\textbf{positive}} & \multicolumn{2}{c}{\textbf{negative}} \\
    \midrule
    \textit{word} & \textit{frequency} & \textit{word} & \textit{frequency} \\
    \midrule
    buy   & 0.0344 & downgraded & 0.0252 \\
    upgraded & 0.0187 & cut   & 0.0240 \\
    deal  & 0.0172 & fly   & 0.0173 \\
    said  & 0.0172 & bank  & 0.0142 \\
    raised & 0.0158 & misses & 0.0110 \\
    fly   & 0.0131 & falls & 0.0106 \\
    order & 0.0107 & neutral & 0.0092 \\
    talks & 0.0094 & hold  & 0.0092 \\
    gets  & 0.0075 & cuts  & 0.0088 \\
    neutral & 0.0071 & sell  & 0.0086 \\
    raises & 0.0065 & buy   & 0.0078 \\
    wins  & 0.0054 & miss  & 0.0078 \\
    hold  & 0.0054 & estimates & 0.0073 \\
    billion & 0.0053 & outlook & 0.0061 \\
    street & 0.0052 & sales & 0.0058 \\
    stake & 0.0048 & profit & 0.0054 \\
    unit  & 0.0047 & earnings & 0.0054 \\
    insider & 0.0046 & tradegate & 0.0050 \\
    buyback & 0.0041 & says  & 0.0049 \\
    outlook & 0.0038 & sees  & 0.0043 \\
    offer & 0.0037 & downgrades & 0.0036 \\
    agrees & 0.0036 & shares & 0.0036 \\
    buys  & 0.0035 & revenue & 0.0035 \\
    says  & 0.0034 & underperform & 0.0029 \\
    group & 0.0032 & underweight & 0.0029 \\
    berenberg & 0.0030 & credit & 0.0028 \\
    near  & 0.0029 & close & 0.0028 \\
    sell  & 0.0029 & lower & 0.0027 \\
    bid   & 0.0026 & results & 0.0027 \\
    tradegate & 0.0026 & loss  & 0.0025 \\
    approval & 0.0025 & leave & 0.0025 \\
    new   & 0.0025 & forecast & 0.0023 \\
    buyout & 0.0025 & guidance & 0.0022 \\
    bank  & 0.0025 & growth & 0.0022 \\
    acquire & 0.0025 & overweight & 0.0021 \\
    outperform & 0.0024 & negative & 0.0020 \\
    rises & 0.0023 & downgrade & 0.0020 \\
    worth & 0.0023 & outperform & 0.0020 \\
    eu    & 0.0022 & loses & 0.0019 \\
    contract & 0.0022 & weight & 0.0019 \\
    close & 0.0021 & seeking & 0.0019 \\
    overweight & 0.0020 & drops & 0.0018 \\
    credit & 0.0020 & equal & 0.0018 \\
    set   & 0.0019 & close & 0.0017 \\
    fiat  & 0.0018 & new   & 0.0017 \\
    gains & 0.0018 & perform & 0.0017 \\
    sees  & 0.0017 & indicated & 0.0016 \\
    merger & 0.0017 & price & 0.0016 \\
    takeover & 0.0016 & target & 0.0016 \\
    \bottomrule
    \end{tabular}}
  \label{tab:addlabel}%
\end{center}
\end{table}%

\end{document}